\documentclass[journal=nalefd,manuscript=letter]{achemso}
\usepackage[T1]{fontenc}

\makeatother

\usepackage{graphicx}
\usepackage{float}
\usepackage{dcolumn}
\usepackage{bm}
\usepackage{hyperref}
\usepackage{chemformula}
\usepackage{siunitx}
\usepackage{verbatim}
\usepackage[T1]{fontenc}
\usepackage[utf8]{inputenc}
\usepackage[symbol]{footmisc}
\usepackage{natbib}
\usepackage{times}
\usepackage{verbatim}
\usepackage{sidecap}
\sidecaptionvpos{figure}{t}
\usepackage[version=4]{mhchem}

\author{Adrian Ebert}

\affiliation{Institute for Quantum Materials and Technologies, Karlsruhe Institute of Technology (KIT), D-76344 Eggenstein-Leopoldshafen, Germany}
\author{Simon Fromme}
\affiliation{Institute for Quantum Materials and Technologies, Karlsruhe Institute of Technology (KIT), D-76344 Eggenstein-Leopoldshafen, Germany}
\author{Lisa Burgert}
\affiliation{Institute for Quantum Materials and Technologies, Karlsruhe Institute of Technology (KIT), D-76344 Eggenstein-Leopoldshafen, Germany}
\author{Umar Rashid}
\affiliation{Institute for Quantum Materials and Technologies, Karlsruhe Institute of Technology (KIT), D-76344 Eggenstein-Leopoldshafen, Germany}
\author{Lukas Gerhard}
\affiliation{Institute for Quantum Materials and Technologies, Karlsruhe Institute of Technology (KIT), D-76344 Eggenstein-Leopoldshafen, Germany}
\email{Lukas.Gerhard@kit.edu}
\author{Julia Feye}
\affiliation{Institute for Inorganic Chemistry, Karlsruhe Institute of Technology, Karlsruhe, Germany}

\author{Senthil Kumar Kuppusamy}
\affiliation{Institute for Quantum Materials and Technologies, Karlsruhe Institute of Technology (KIT), D-76344 Eggenstein-Leopoldshafen, Germany}

\author{Barbora Brachnakova}
\affiliation{Institute for Quantum Materials and Technologies, Karlsruhe Institute of Technology (KIT), D-76344 Eggenstein-Leopoldshafen, Germany}

\author{Timo Neumann}
\affiliation{Institute of Inorganic Chemistry, University of Tübingen, Tübingen, Germany}

\author{Mario Ruben}
\affiliation{Institute of Nanotechnology, Karlsruhe Institute of Technology (KIT), D-76344 Eggenstein-Leopoldshafen, Germany}
\alsoaffiliation{Institute for Quantum Materials and Technologies, Karlsruhe Institute of Technology (KIT), D-76344 Eggenstein-Leopoldshafen, Germany}
\alsoaffiliation{Centre Européen de Sciences Quantiques (CESQ), Institut de Science et d’Ingénierie Supramoléculaires (ISIS), BP 70028, 67083, Strasbourg Cedex, France.}

\author{Peter W. Roesky}
\affiliation{Institute for Inorganic Chemistry, Karlsruhe Institute of Technology, Karlsruhe, Germany}

\author{Michael Seitz}
\affiliation{Institute of Inorganic Chemistry, University of Tübingen, Tübingen, Germany}

\author{Wulf Wulfhekel}
\affiliation{Institute for Quantum Materials and Technologies, Karlsruhe Institute of Technology (KIT), D-76344 Eggenstein-Leopoldshafen, Germany}

\title{Photophysical properties of Eu$^{3+}$ complexes approaching electronic contact to a metal surface}
\usepackage{xspace}

\usepackage{upgreek}
\usepackage{miller}

\usepackage{placeins}

\newcommand{\tta}{\text{[Eu(tta)$_3$(bpy)]}\,}

\begin{document}
\clearpage
\begin{abstract}
The application of rare-earth complexes in electrically driven light sources poses a series of challenges that require specific optimization of the molecular photophysical properties.
Here, we present a report on films of three different Eu$^{3+}$ complexes characterized in terms of emission spectra and fluorescence decay. We compare molecular complexes in powder form and sublimed films, in films on glass and on a metal surface, and in films of thicknesses down to less than 3 nm (< 3 ML), approaching electrical coupling. Our photoluminescence experiments supported by scanning tunneling microscopy of sub-monolayers indicate that Eu$^{3+}$ trensal-complexes are less affected by sublimation and more stable on the metal surface than typical beta diketonate complexes, making them promising candidates for electroluminescence devices.
\end{abstract}

\section{Introduction}
     Miniaturization of electronic devices with the ultimate goal of electrically controlled single-molecule devices by exploiting intrinsic quantum properties requires a comprehensive understanding of the wide range of effects that occur in the presence of a metal. \sloppy
     Rare-earth complexes appear to be ideally suited for nanoscale opto-electronic devices, based on their high ligand-based absorption and strong rare-earth ion luminescence, sharp and environment-insensitive emission spectra, and rich photophysics \cite{kuppusamy_spin-bearing_2024, song_smart_2025}.  
     Functional electronic devices intrinsically require electrodes to make electrical contact with the organic moiety.
     For light in the visible range, typical electrode metals such as Ag act as a mirror changing the photonic mode density of a nearby emitter, which has been the subject of research since the pioneering experiments of Drexhage\cite{DREXHAGE1970693, Noginova2009, worthing_modification_1999, noginova_modification_2012, Chance1978, Kuhn1970, Morawitz1969}. Depending on the structure and nature of the metal electrode, coupling of molecular excitons to plasmons leads to quenching \cite{amos_modification_1997, Pockrand1980, Chance1978, Ford1984} or enhancement of the light emitted in the far field \cite{MOR2025, Khan20215532, LIANG2017181, Pang2015527, Schmidt202417109}.
     Electroluminescence\cite{10.1063/1.2229577}, which refers to the luminescence of electrically excited molecules, intrinsically requires some degree of electrical contact to an electrode, which leads to additional potential de-excitation by tunneling of charge carriers at very small distances to the electrodes.
     Fluorescence of films composed of organic chromophores has been explored at small distances from metal surfaces in the past \cite{campion_electronic_1980,kuhnke_c_1997} and no indication for the onset of exciton quenching via tunneling has been found for distances down to about 1 nm. However, it is not straightforward to transfer these results to radiating 4f excitations of lanthanide (Ln) ions: On the one hand, the vastly different radiative lifetimes ($\approx 10^{-9}$ s in typical chromophores  vs $\approx 10^{-3}$ s in Ln ions) suggest that drastically enhanced decoupling from the metal surface would be required. On the other hand, the localized nature of the 4f excitations already significantly isolates the excitations from the environment. In addition, the small cross-section of 4f excitations typically makes excitation via antenna ligands and an efficient intra-molecular energy transfer to the Ln ion necessary. 
     Besides these direct optical and electrical effects on the light emission, a metal surface also affects growth of adsorbed molecular complexes and even facilitates fragmentation \cite{knaak_fragmentation_2019}. Finally, rare-earth complexes are prone to bleaching under repeated optical excitation, an effect that has recently been linked to reduction \cite{melo_unmasking_2025} and that might be affected by the electron bath of a nearby metal as well.
     
     This calls for a systematic study of the optical properties of Ln complexes in very close proximity to a metal surface under well-controlled conditions. 

     \begin{figure}
        \centering
        \includegraphics[width=\linewidth]{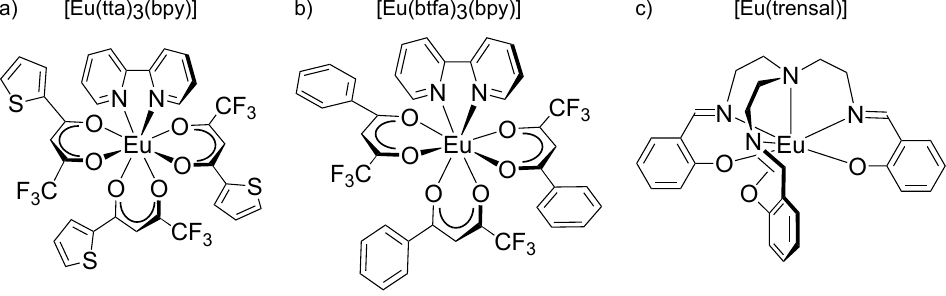}
        \caption{Structure representation of [Eu(tta)$_3$(bpy)], [Eu(btfa)$_3$(bpy)] and [Eu(trensal)] complexes.}
        \label{all complexes ink}
    \end{figure}
     
     Here, we report on characterization of the thin film growth of [Eu(tta)$_3$(bpy)] \cite{CHITNIS2019102302,CHITNIS201761,GAMEIRO2002385,Liu2013LDS,Ebert} (Fig.\,\ref{all complexes ink} (a)), [Eu(btfa)$_3$(bpy)] \cite{Batista1998,GAMEIRO2002385,DESA199823,QUIRINO2011964,Ebert} (Fig.\,\ref{all complexes ink} (b)) and [Eu(trensal)] \cite{kuppusamy_observation_2023} (Fig.\,\ref{all complexes ink} (c)) complexes on glass and Ag substrates  under ultra-high vacuum conditions. Photoluminescence spectra (PL) and time-correlated single photon counting (TCSPC) of thin films down to about 3 monolayers (ML) show evidence of strong PL quenching of luminescence due to electromagnetic coupling to the flat metal surface and significant bleaching in the absence of oxygen at 4.5 K. Both PL and scanning tunneling microscopy (STM) of sub-monolayers indicate that [Eu(trensal)] is less affected by sublimation than the two beta diketonate complexes \cite{stoll_magnetic_2016}.

\FloatBarrier\section{Experimental}

\FloatBarrier\subsection{Optical set-up}

     Low-temperature STM measurements were performed using a custom-built ultra-high vacuum (UHV) STM operating at 4.5\,K and a base pressure of approximately $10^{-10}$\,mbar~\cite{edelmann_light_2018}. 
    
\begin{figure}
        \centering
        \includegraphics[width=0.5\linewidth]{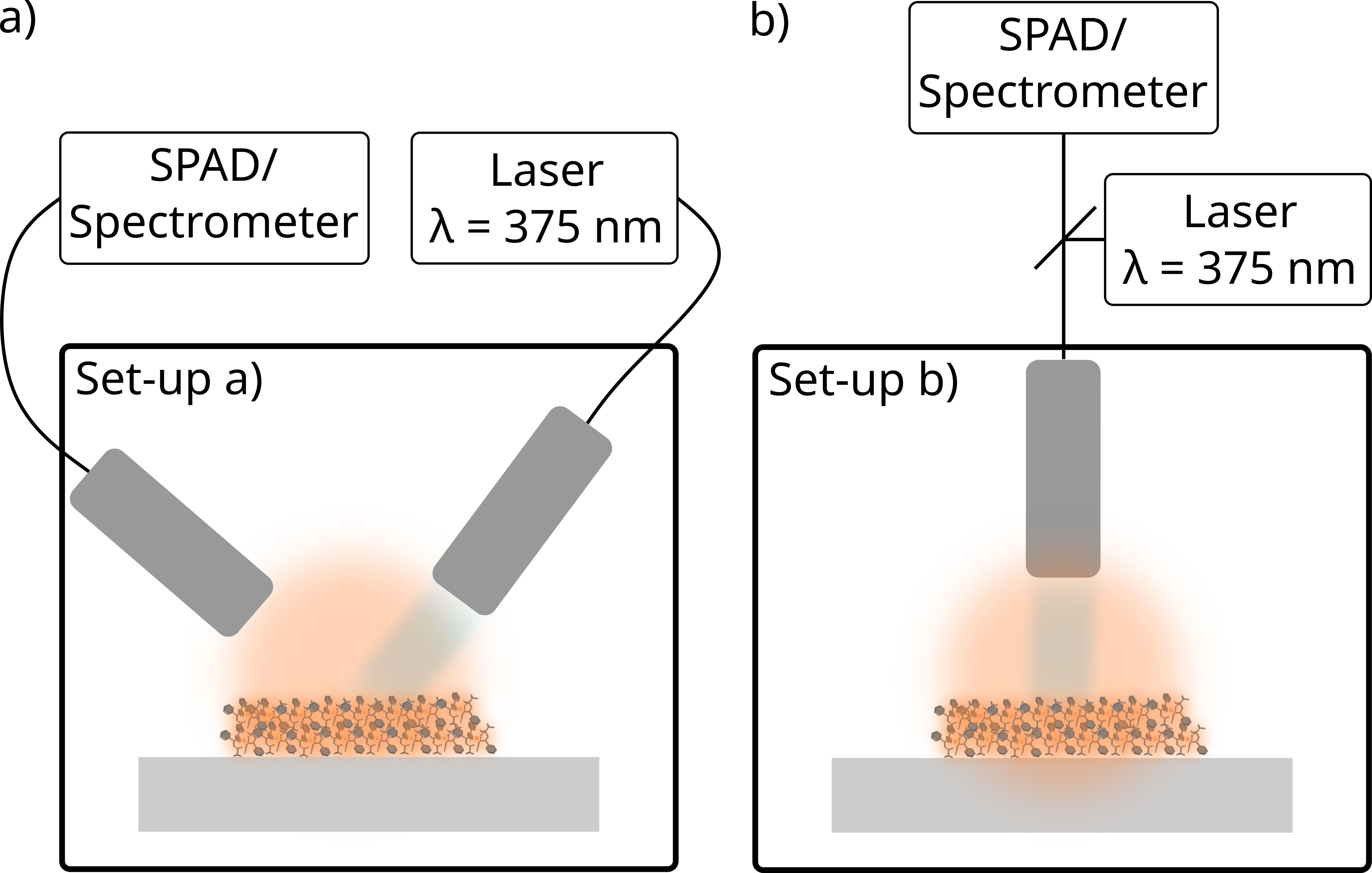}
        \caption{Schematic drawings of the photoluminescence (PL) set-ups. a) Two optical fibers positioned at an angle and about 0.5 mm from the sample for excitation and collection of the PL response.
        b) Single optical fiber positioned perpendicular to the sample at a distance of about 30 $\upmu$m. 
        }
        \label{optical set-up}
    \end{figure}

     PL spectra were acquired using a spectrometer with a focal length of 150\,mm and a diffraction grating of 300\,grooves/mm. The maximum spectral resolution of the system is approximately 2\,nm with a slit width of $10\,\upmu$m, but for most spectra presented here spectral resolution has been traded for detection efficiency. All photon spectra presented are corrected for the wavelength-dependent collection efficiency of the detection setup. Excited-state lifetimes were determined via TCSPC using a single photon avalanche diode (SPAD) for detection with bandpass filter of 610 $\pm$ 10 nm. Two set-ups were used for PL measurements, both with sample and optical fibers in UHV conditions. Set-up a) (Fig.\,\ref{optical set-up} (a)) features two multimode fibers (diameter 200 $\upmu$m) positioned at an angle about half a mm away from the sample. One optical fiber guides the light for excitation towards the sample at room temperature (RT), the other one collects the PL response and guides it towards the SPAD for TCSPC measurements, or towards the spectrometer. The plane defined by the two optical fibers is tilted with respect to the sample to minimize collection of reflected light. Set-up b) (Fig.\,\ref{optical set-up} (b)) features only one multimode fiber (diameter 200 $\upmu$m) positioned perpendicular to the sample at a distance of about 30 $\upmu$m inside the cryostat at a temperature of 4.5\,K. 

\subsection{Sample preparation}
    The complexes were synthesized according to \cite{CHITNIS2017,Batista1998}.
   Ag(111) single-crystal substrate was prepared by repeated cycles of argon ion sputtering (1.5\,keV), followed by annealing at temperatures between 500 and $600\,^{\circ}$C.
    Thermal sublimation of [Eu(btfa)$_3$(bpy)], [Eu(tta)$_3$(bpy)], and [Eu(trensal)] complexes was carried out using a Kentax three-cell evaporator. The areal densities of the sublimed films were calibrated using a quartz microbalance and their thickness was estimated using the volume densities from the literature \cite{seward_dimeric_2001}. Molar mass allows to estimate The thicknesses in molecules per nm$^2$ can be calculated by dividing the areal density by the molar mass.

\section{Results and discussion}

\subsection{Comparison of photophysical properties of powder and thin films of \break [Eu(btfa)$_3$(bpy)], [Eu(tta)$_3$(bpy)] and [Eu(trensal)]}

\begin{figure}[h]
        \centering
        \includegraphics[width=0.5\linewidth]{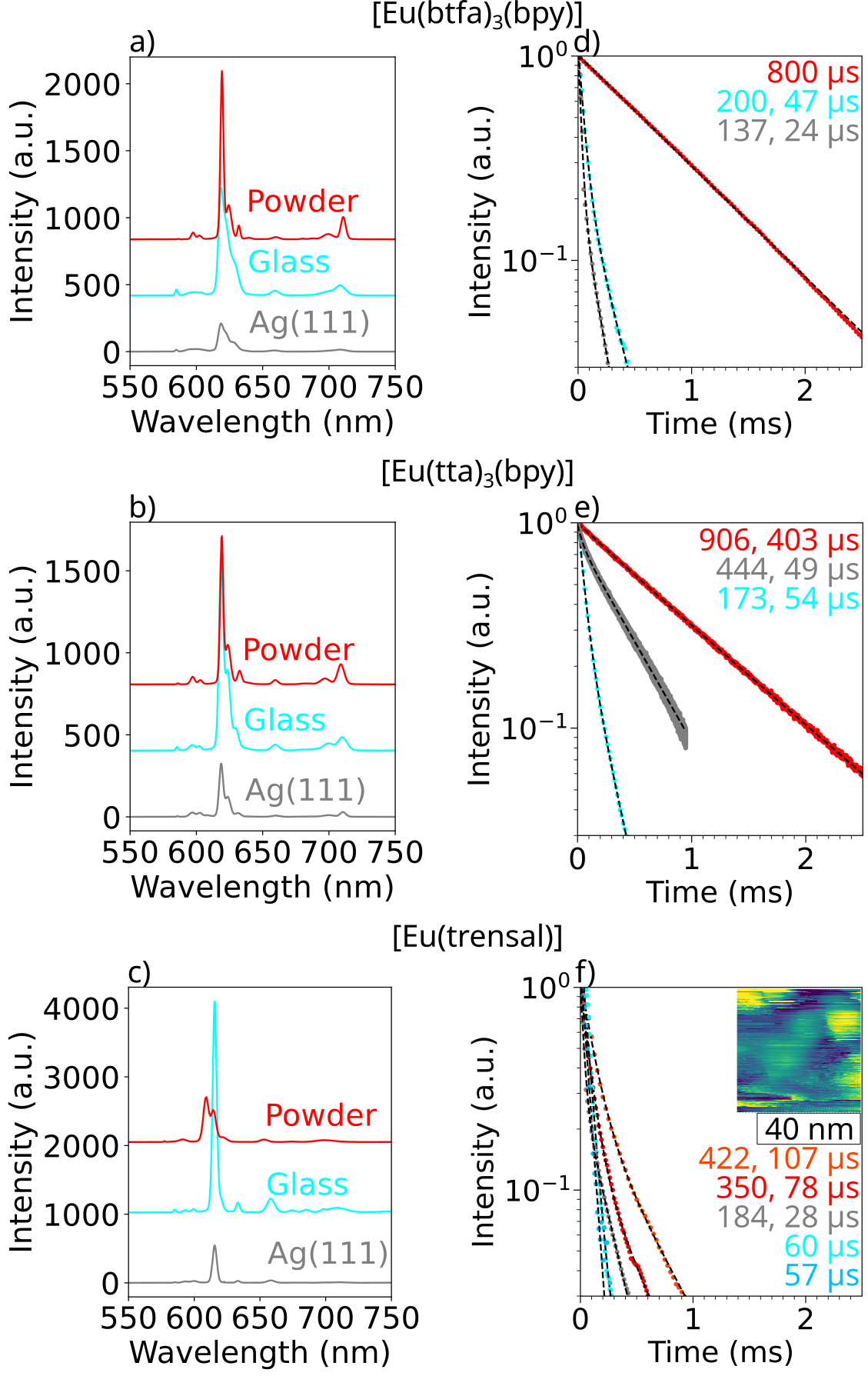}
        \caption{Photophysical properties of thin films of [Eu(btfa)$_3$(bpy)], [Eu(tta)$_3$(bpy)] and [Eu(trensal)] under excitation with $\lambda=375$ nm. a)-c) PL spectra of powder and films sublimed onto glass and Ag(111) with a density of  $\sim$ 22.5 molecules/nm$^2$ at RT. The intensity of the PL spectra has been normalized to the integrated intensity of the $^5D_0\rightarrow\,^7F_1$ transition. Graphs shifted for clarity. d)-f) Decay of the intensity of the $^5D_0 - ^7F_2$ transition (recorded using a bandpass filter of 610 $\pm$ 10 nm) in powder and thin films on glass and Ag samples. The $\tau_{obs}$ obtained by fitting the sum of two exponential decays are included. The colors correspond to the different samples; red, orange: powder; cyan, blue: 22.5 molecules/nm$^2$ film on glass at RT, gray: 22.5 molecules/nm$^2$ film on Ag(111) at 4.5 K. The inset in f) shows an STM topography image of a [Eu(trensal)] thin film on Ag(111) scanned at -10 V, 1 pA, indicating relatively stable conductivity.}
        \label{spectra glass ag}
    \end{figure}

    [Eu(btfa)$_3$(bpy)], [Eu(tta)$_3$(bpy)] and [Eu(trensal)] complexes were sublimed with a thickness of $\sim$ 22.5 molecules/nm$^2$ ($\approx 21$ nm in case of \tta) on glass (microscope slide) and Ag(111) substrates in UHV. Excitation with $\lambda=$ 375 nm results in the emission spectra shown in the left panels of Fig.\,\ref{spectra glass ag} with the well known series of emission bands corresponding to transitions from the excited $^5D_0$ to the $7F_{J=0-4}$ ground states of the Eu$^{3+}$ ion \cite{BINNEMANS20151}. All PL spectra were normalized to the magnetic transition $D_0\rightarrow F_1$. The right panels display the corresponding time evolution of the PL intensity of the main peak corresponding to the $^5D_0 \rightarrow\, ^7F_2$ transition. This peak is dominant for all samples, giving the Eu$^{3+}$ emission its characteristic red color. 

    To discuss the variations of the emitted light, we note that four different mechanisms affect the PL spectra and fluorescence lifetimes. First, there can be chemical changes \cite{horrocks_lanthanide_1979} or differences in the molecular arrangement in the three types of samples affecting the emission spectra. Second, the emission characteristics are affected by the optical environment of the molecules. In powder, it is isotropic, the vacuum-glass interface provides a discontinuity in the refractive index, and the metallic Ag substrate is reflective \cite{kunz_changes_1980, Noginova2009}. Third, on the metallic substrate surface plasmons may interact with the molecules \cite{Pockrand1980}. Finally, emission can be quenched by electronic hybridization of 4f states and the conduction electrons of the metal substrate.

    The highest energy $^5D_0\rightarrow\,^7F_0$ transition is strictly forbidden according to the standard Judd-Ofelt theory but can be induced by $J$ mixing caused by the ligand field \cite{BINNEMANS20151,Porcher1980207, Tanaka199416917, Tanaka19954171, Malta198165,CHEN2005419}. This emission band can easily be seen in the sublimed samples of [Eu(btfa)$_3$(bpy)] and [Eu(tta)$_3$(bpy)] on both substrates while it is nearly absent in the powder sample. This indicates a modified ligand field effect in the sublimed samples, which might be explained by a deformation due to a different inter-molecular arrangement in the film or even the cleavage of one of the ligands by sublimation \cite{stoll_magnetic_2016}.

    This is supported by the observation that the individual stark levels of the $^5D_0\rightarrow\,^7F_2$ and the $^5D_0\rightarrow\,^7F_4$ bands are sharper in the emission of [Eu(btfa)$_3$(bpy)] and [Eu(tta)$_3$(bpy)] in powder compared to the sublimed films, while they are similar for the [Eu(trensal)] samples (see left panels of Fig.\,\ref{spectra glass ag}).
    
    To qualitatively compare the dynamics of the light emitting excited states of the three different complexes, we measured the  decay of the intensity of the $^5D_0 \rightarrow\,^7F_2$ transition (bandpass filter of 610 $\pm$ 10 nm) after prolonged illumination (0.1 to 0.5 ms, $\lambda=$ 375 nm) as shown in Figure \ref{spectra glass ag} (d)-(f). The experimental data deviates from a single exponential decay in particular in the sublimed films, which is why a fit with a sum of two exponential decays is shown. This deviation indicates that Eu ions experience different local environments, possibly depending on the distances to the underlying substrate. 

    Comparing the overall emission dynamics of the three complexes, [Eu(trensal)] shows about a factor of two shorter lifetimes in powder form ($\approx 800\, \upmu$s vs $\approx 400\, \upmu$s) and also mostly in the films (see Fig.\ \ref{spectra glass ag} (d)-(f)), in agreement with previous work \cite{BORGES2016654, Ebert, kuppusamy_observation_2023}. For [Eu(tta)$_3$(bpy)] and [Eu(btfa)$_3$(bpy)], the observed lifetime $\tau_{\mathrm{obs}}$ in thin sublimed films is clearly shorter than in powder, in agreement with our previous work and early literature \cite{Ebert,kunz_changes_1980} (see Fig.\ \ref{spectra glass ag} (d),(e)), while this effect is less pronounced for [Eu(trensal)].
    Together with the changes in the PL spectra, this indicates that sublimation of [Eu(btfa)$_3$(bpy)] and [Eu(tta)$_3$(bpy)] which have C1 symmetry, has more pronounced consequences in terms of lack of crystallinity and integrity of the complexes. The [Eu(trensal)] ligand structure with C3 symmetry seems to be more robust compared to beta diketonates [Eu(tta)$_3$(bpy)] and [Eu(btfa)$_3$(bpy)]. 

    In addition to reduced symmetry in the sublimed films, hydroxyl groups on the glass surface can lead to quenching of nearby emitters via phonons.

    In addition to the increased structural disorder in the sublimed film, increased refractive index of the surrounding medium \cite{BINNEMANS20151} (glass vs. air) and the discontinuous jump in the refractive index modifying the photonic density of states at the interface may be reasons for increased radiative decay for films on glass. The controlled growth of thin films on glass is beyond the scope of this paper and we therefore refrain from quantitative analysis of the observed decay rates (estimation of radiative rates based on Judd-Ofelt analysis).
    
    The magnetic dipole transition $^5D_0\rightarrow\,^7F_1$ is typically assumed to be less sensitive to the local environment, and thus the corresponding radiative decay rate is approximately constant for all Eu$^{3+}$ complexes \cite{BINNEMANS20151}. Based on this assumption, the relative intensities of the other transitions are proportional to the corresponding radiative decay rates \cite{BINNEMANS20151, Görller-Walrand19913099}.

    Note that on the metal surface, the modes of the electromagnetic field with a wave vector normal to the surface display a node in the electric field and a maximum in the magnetic field near the interface due to boundary conditions. As our setup (see Fig.\ \ref{optical set-up}) mainly detects light emitted normal to the surface, this property leads to a relative suppression of the electric dipole emission normal to the surface compared to that of the magnetic dipole emission. When the spectra are normalized to the magnetic transition intensity, this effect manifests as a relative decrease of the measured electric dipole peaks as observed in all spectra on Ag(111). 

     The Ag surface leads to dissipation of molecular excitations in its vicinity via plasmon excitation depending on the orientation of the transition dipole moment with respect to the substrate, the distance to the substrate and the geometry of the substrate \cite{PHILPOTT1972,Hussain2014, Novotny2012,Noginova2009,LAKOWICZ2002, doi:10.1021/acsphotonics.5b00397, Pockrand1980}, which leads to a rather large variation within nominally identical films. 
     A more detailed study of the influence of [Eu(tta)$_3$(bpy)] film thickness on the fluorescence decay is discussed below.

     Hybridization of the 4f states of Eu$^{3+}$ and the metal substrate that allows charge transport with time constants comparable to the PL lifetime would drastically suppress photon emission, which we have not observed, despite a significant conductance across molecular layers as shown by the STM topography in Figure \ref{spectra glass ag} f).

   \begin{SCfigure}
        \centering
        \includegraphics[width=0.45\linewidth]{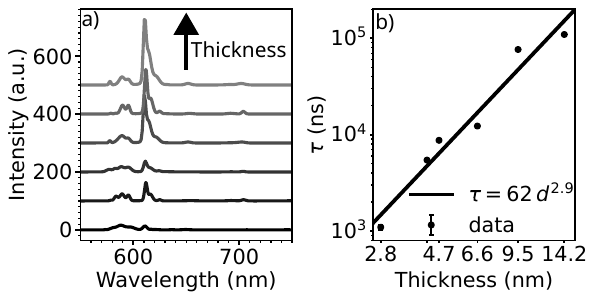}
        \caption{a) Pl spectra of [Eu(tta)$_3$(bpy)] films with different thicknesses from 2.8 nm to 14.1 nm on Ag(111) at 4.5 K. The intensity of the PL spectra has been normalized to the integrated intensity of the $^5D_0 - ^7F_1$ transition. 
        Graphs shifted by 100 a.u. for clarity. b) Lifetimes from single exponential fitted to the fluorescence decays of the different film thicknesses (excitation via UV: 375 nm, single pulse of 30 ps).}
        \label{Thickness dependency Ag111 4K spectra}
    \end{SCfigure}

\subsection{Photophysical properties of [Eu(tta)$_3$(bpy)] as a function of its film thickness}    
\begin{SCfigure}
        
        \includegraphics[width=0.45\linewidth]{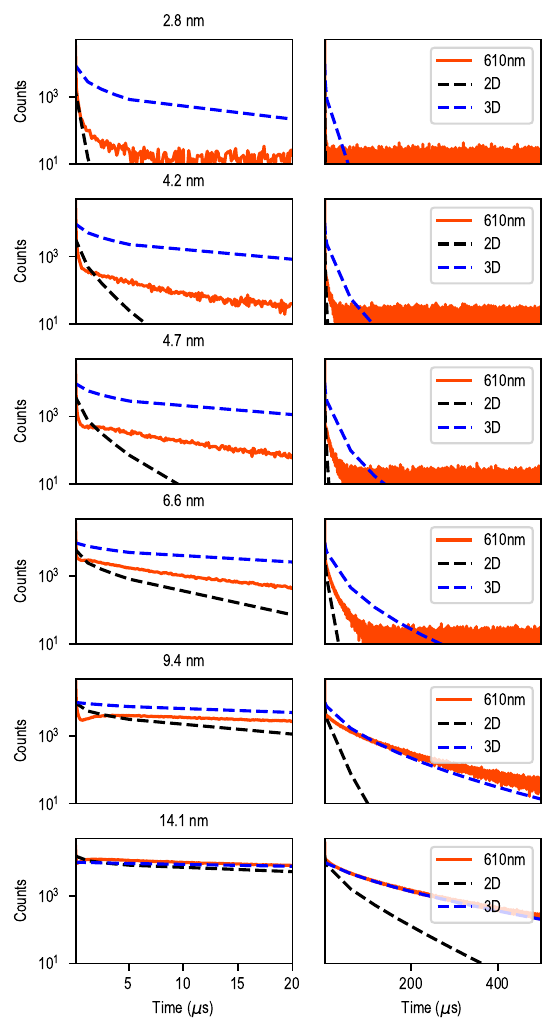}
        \caption{TCSPC (red) recorded for [Eu(tta)$_3$(bpy)] films with different thicknesses on Ag(111) at 4.5 K using a bandpass filter of 610$\pm$10 nm (excitation via UV: 375 nm, single pulse of 30 ps). Left and right columns show the same data on different time ranges. Dashed black/blue lines show the decay estimated from distance dependent dissipation of dipole radiation as discussed in the text for perfect 2D growth / poisson height distribution (3D growth), respectively, based the nominal thickness of each film and a single overall intensity factor. Note that no fitting to the thickness dependent decay has been performed here. }
        \label{Thickness dependency Ag111 4K decay}
    \end{SCfigure}

    To quantitatively investigate the impact of the metal substrate on light emission, films of [Eu(tta)$_3$(bpy)] with densities between 3 and 15 molecules $\cdot$ $\mathrm{nm^{-2}}$ were sublimed on a flat Ag(111) single crystal. \sloppy
    Here we were careful to keep the exact same conditions for the film growth for different thicknesses in order to allow a direct comparison. Assuming a density of 1.7 $\mathrm{g\cdot cm^{-3}}$ \ \ \cite{seward_dimeric_2001} the areal density calibrated by the quartz balance translates to thicknesses of 2.8 to 14.1 nm. PL spectra and TCSPC were measured at 4.5 K with an optical fiber placed $\approx 30 \, \upmu$m above the film (see Fig.\,\ref{optical set-up} (b)). 
    As discussed above, the different boundary conditions of the electric and magnetic components of the electromagnetic field at the metal surface in combination with our detection setup perpendicular to the surface favor MD emission over ED emission. This effect is strongly enhanced for thinner films. For the thinnest film, the MD $^5D_0 \rightarrow\, ^7F_1$ transition is even more intense than the ED $^5D_0 \rightarrow\, ^7F_2$ transition (see Fig.\ \ref{Thickness dependency Ag111 4K spectra}(a)). In addition, the overall integrated intensity per molecule of the emission spectra drastically decreases with decreasing film thickness. The TCSPC measurements, performed for each film thickness, show that the decay rates drastically increase with decreasing film thickness (see Fig.\ \ref{Thickness dependency Ag111 4K decay}). If approximated by a single exponential decay, the observed decay rate scales as $\gamma_{\mathrm{obs}} \propto d^{-2.9}$ (see Fig.\ \ref{Thickness dependency Ag111 4K spectra} (b)).

    This is well expected for radiation being dissipated into the nearby metal in form of plasmons. At short distances, the dissipation rate $\gamma_{\text{quench}}$ of dipolar emission in the metal bulk is proportional to the radiative rate in free space $\gamma_{\text{rad0}}$ and the distance $d$ to the power 3 between the molecular emitter and the metal surface \cite{Ford1984, bharadwaj_spectral_2007, faggiani_quenching_2015,CAMPION1980447}. 
    \begin{equation}
    \begin{split}
    \gamma_{\text{quench}} = \gamma_{\text{rad0}} \cdot \frac{3}{16k^3} \, \text{Im}\!\left( \frac{\varepsilon(\omega) - \varepsilon(\mathrm{mol})}{\varepsilon(\omega) + \varepsilon(\mathrm{mol})} \right) \cdot \frac{p_x^2 + p_y^2 + 2p_z^2}{|\mathbf{p}|^2} \cdot \frac{1}{d^3} 
    \\
    \approx \gamma_{\text{rad0}} \cdot \frac{4030}{(d [\mathrm{nm}])^3}
    \end{split}
    \label{d_cube}
    \end{equation}

    Here, we approximated $\lambda = 615$ nm (corresponding to the $^5D_0 \rightarrow\, ^7F_2$ transition), $\varepsilon(\mathrm{mol})$=2.56 \cite{BINNEMANS20151} and $\varepsilon(\omega)$ = -16+0.6\textit{i} \cite{yang_optical_2015} denote the permittivities of the metal surface and the molecular film, respectively. A random orientation of the transition dipoles results in $\frac{p_x^2 + p_y^2 + 2p_z^2}{|\mathbf{p}|^2} = 4/3$. In this way, the decay can be estimated from literature values as $\gamma_{\mathrm{obs}} = \gamma_{\mathrm{rad0}}*(1+ \frac{4030}{(d [nm])^3})+ \gamma_{\mathrm{nonrad}}$. Note that the energy transferred to the substrate in the form of plasmonic excitations is unlikely to be radiated to the far field for the present case, because of the momentum mismatch between the plasmomic modes of the atomically flat Ag(111) that we chose for simplicity and the radiative modes in the far field.
    As the film growth mode is unknown, we compare our experimental data to two scenarios: a perfect 2D growth with vanishing roughness of the film and a Poisson distribution of molecules (3D growth). In this way, we can model the decay of each film as a superposition of individual layers with the corresponding distance dependent decay rate. This leaves only a global factor as a free parameter that includes excitation/detection efficiency and which amounts to a constant offset in the logarithmic plots. As is shown in Figure \ref{Thickness dependency Ag111 4K decay}, the experimental data indeed falls right in between the extreme scenarios of 2D and 3D growth for all thicknesses. This indicates no significant additional quenching pathways e.g. via electron transport. 

\begin{SCfigure}
        \centering
        \includegraphics[width=0.45\linewidth]{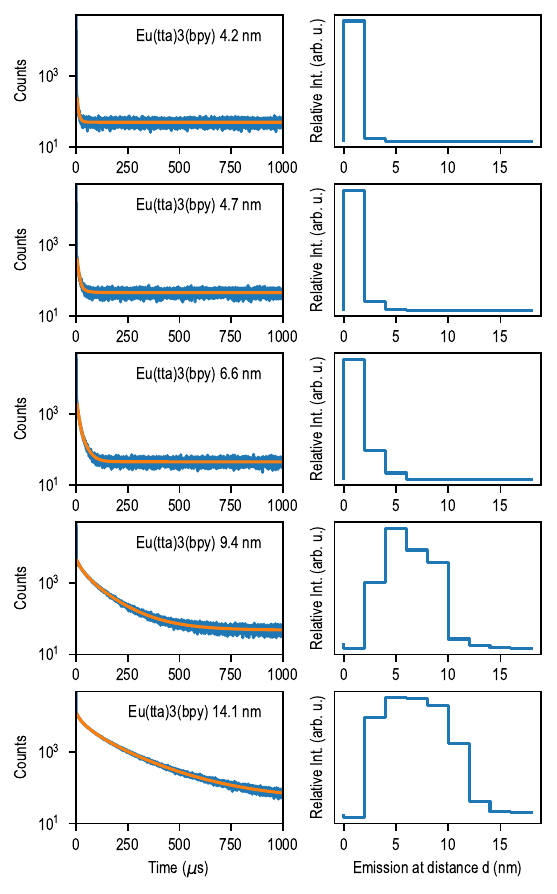}
        \caption{Experimental TCSPC data fitted with a sum of exponential decays (left panels) and corresponding thickness distribution extracted from the fit (right panels). The fit reveals the relative contribution from each layer.}
        \label{fig:LaplaceTransformOfDecay}
    \end{SCfigure}

    If the molecular decay constants are known in bulk, equation \ref{d_cube} can be used to estimate the distance of a emitter from the surface from its PL decay. In a film, the distribution of the distances of the emitters from the surface can be estimated by fitting a sum of exponentials with decay constants corresponding to equally spaced distances to the experimental decay. Here, we use d = 0,2,4,..., 20 nm as a basis set (see Fig. \ref{fig:LaplaceTransformOfDecay}). The resulting estimates for the distance distribution of the emitters agree well with the expected thickness distributions of the film with the nominal thickness determined by the quartz microbalance, in particular for the thicker films.

\subsection{Decrease of fluorescence intensity under illumination}  
    \begin{SCfigure}
        \centering
        \includegraphics[width=0.45\linewidth]{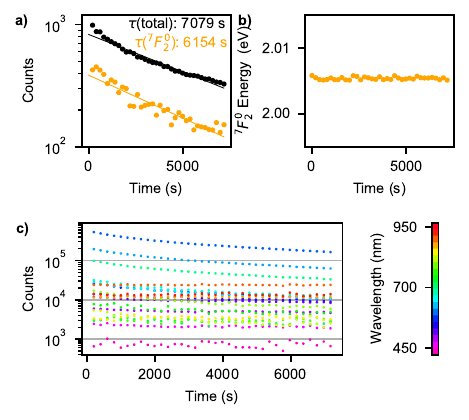}
        \caption{Decrease of photoluminescence intensity of a film of [Eu(tta)$_3$(bpy)] on Ag(111) with a density of  $\sim$ 22.5 molecules/nm$^2 \approx 21$ nm with time at 4.5 K under illumination (375 nm). a) Fluorescence intensity as a function of time integrated over all emission bands (black dots) and the most prominent $^5D_0 \rightarrow\, ^7F_2$ band (yellow dots). Exponential fits shown by solid lines. b) Peak position of the most prominent $^5D_0 \rightarrow\, ^7F_2$ band. c) Fluorescence intensity decreases at all wavelengths. The range from about 450 nm to about 950 nm has been divided into 20 equal bins.}
        \label{Bleaching_Oview}
    \end{SCfigure}

    Bleaching by exposure to UV light is a well documented effect for Eu$^{3+}$ complexes. Most studies focus on complexes in solution \cite{B506915G}, embedded in host matrices\cite{JIMENEZ2018271, PAGNOT2000572, GAMEIRO2001820, XU2000351} or single crystals\cite{Lima2013} and are limited to qualitative conclusions. Since our ultimate aim is to investigate the luminescence of electrically contacted single molecules, we quantified the photo-degradation process of a [Eu(tta)$_3$(bpy)] film with a density of around 22.5 molecules/nm$^2$ on Ag(111) under exposure to UV light with $\lambda=375$ nm in UHV at T = 4.5 K. The PL spectrum was repeatedly recorded using the setup shown in Figure \ref{optical set-up} (b). Figure\,\ref{Bleaching_Oview} (a) shows that the intensity integrated over the whole spectral range and the intensity of the $^5D_0 - ^7F_2$ transition decrease significantly over time. The photo-bleaching lifetime of the [Eu(tta)$_3$(bpy)] is estimated to be $\approx6\cdot10^3$ s from an exponential fit to the time dependent intensity (see Fig.\ \ref{Bleaching_Oview} (a)) and corresponds to an average number of about $10^5$ photons emitted by each Eu complex before the initial intensity of the PL signal drops to 1/e. This photo-bleaching probability of the order of $10^{-5}$ is within the range of typical organic dyes in water and at room temperature ($10^{-3}$ - $10^{-7}$) \cite{eggeling_photobleaching_1998} but too high to perform extended single-molecule experiments. Within the margins of error due to limited wavelength resolution of the setup, there is no change of the emission energy (see Fig. \ref{Bleaching_Oview}b) and no appearance or disappearance of any peak (see Fig.\ \ref{Bleaching_Oview} (c)) which would indicate fragmentation of the complex or reduction of Eu$^{3+}$ to Eu$^{2+}$ that has been previously reported \cite{melo_unmasking_2025}.

\FloatBarrier\subsection{Surface deposition and STM analysis}

    \begin{figure}
        \centering
        \includegraphics[width=0.5\linewidth]{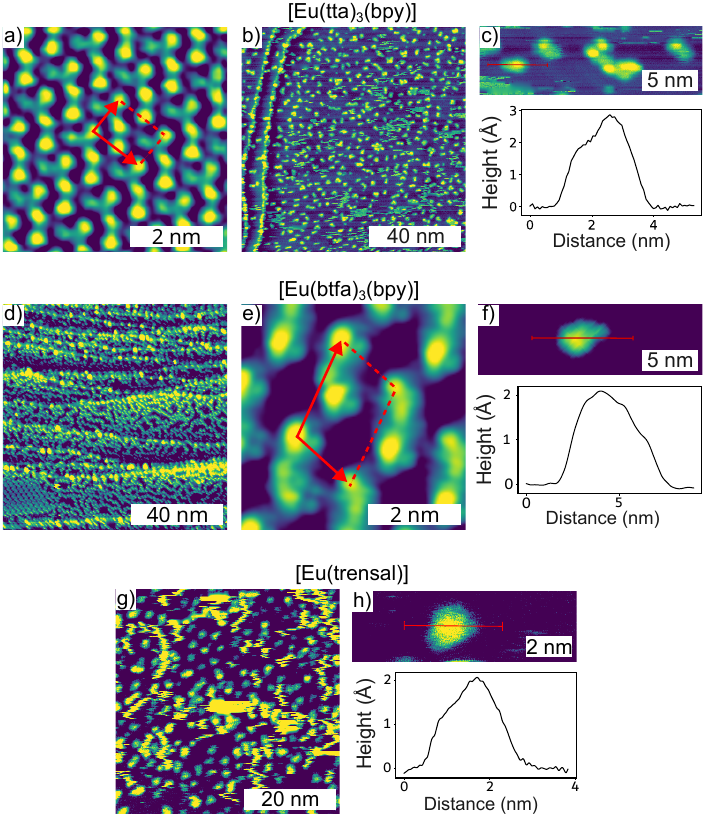}
        \caption{(a) Close-up STM image of lattice formation on a film of [Eu(tta)$_3$(bpy)] on Ag(111) sample with unit cell indicated in red. (2.0 V, 1 pA). (b) STM image of [Eu(tta)$_3$(bpy)] on Au(111) sublimed at 185 °C. (2.0 V, 10 pA). (c) Height profile across island of [Eu(tta)$_3$(bpy)] on Ag(111) sublimed at 185 °C after degassing the crucible at 150 °C for  $\sim$ 10 h. (2.0 V, 10 pA). (d) Large scale STM overview of [Eu(btfa)$_3$(bpy)] close to 1 ML of on Ag(111). (2.0 V, 10 pA) (e) Close-up STM image of lattice formation on a [Eu(btfa)$_3$(bpy)] on Ag(111) sample with unit cell indicated in red. (2.0 V, 10 pA). (f) Height profile across island of [Eu(btfa)$_3$(bpy)] (2.0 V, 1 pA). (g) Large scale STM overview of [Eu(trensal)] on Au(111). (2.0 V, 1 pA). (h) Height profile across a molecule on a [Eu(trensal)] on Au(111) sample with corresponding STM image. (2.0 V, 1 pA).}
        \label{ligand lattice}
    \end{figure}
    The adsorption configuration of deposited molecular complexes is relevant for several reasons. As discussed above, the PL spectra shown in Fig.\ \ref{spectra glass ag} and the corresponding decay experiments suggest a reduced symmetry in the sublimed films in comparison to powder, in case of films of [Eu(tta)$_3$(bpy)] and [Eu(btfa)$_3$(bpy)], but not in case of [Eu(trensal)]. In addition, the orientation of the transition dipole moments influences the coupling of light to the underlying metal substrate and into the detection system. Therefore, it is worth investigating the adsorption and possible self-assembly of the molecules on the substrate with STM topography. We previously reported on the self assembly of Eu complexes\cite{Ebert}, including [Eu(tta)$_3$(bpy)] and [Eu(btfa)$_3$(bpy)].
    Recent sublimation experiments at elevated temperatures (185 °C instead of 150 °C, degassing at 150 °C instead of 100 °C) revealed that sublimation of [Eu(tta)$_3$(bpy)] can result in the same pattern with the same unit cell (see Fig.\,\ref{ligand lattice} (b)) as those previously identified with [Eu(btfa)$_3$(bpy)] \cite{Ebert}. In addition, at a sublimation temperature of 185 °C, the deposition of [Eu(tta)$_3$(bpy)] leads to a higher number of significantly larger objects without any apparent preferred orientation (see Fig.\,\ref{ligand lattice} (e,f)) which are likely to be entire complexes. This leads us to revise our previous conclusion and we now suggest that the self-assembled layers shown in b) and\, \cite{Ebert} are composed of fragments of the [Eu(tta)$_3$(bpy)] and [Eu(btfa)$_3$(bpy)] complexes. Bi-pyridine is part of both complexes and the observed pattern is in good agreement with 2,2,-bipyridine on Au(111)\cite{DRETSCHKOW1998121,Cunha1996}.
    
    Similarly, sublimation of [Eu(btfa)$_3$(bpy)] at 185 °C for 1 min shows a different structure formation than previously reported. The surface in some areas is covered with a lattice  likely consisting of intact [Eu(btfa)$_3$(bpy)] complexes (Fig.\,\ref{ligand lattice} (g),(h)). The unit cell size is 3.73 nm$^2$ with two molecules per unit cell which is about three times larger than what we previously reported \cite{Ebert}. 

    Sublimation of [Eu(trensal)] at a crucible temperature of 270 °C for $\sim$ 9 s onto Au(111) at RT, resulted in a single type of homogeneously distributed molecules of approximately 2 \textbf{\AA} height (Fig.\ \ref{ligand lattice} (i,(j)) without apparent preferred orientation. All three complexes show significant mobility in the STM junction, despite low temperature and low tunneling currents during scanning. Suitable complexes for single-molecule experiments would require increased stability both during the sublimation process and in terms of anchoring to the surface.

\subsection{Conclusion}

We have characterized the optical properties of sublimated thin films of three Eu$^{3+}$ complexes on glass and Ag substrates and of varying thickness. In sublimated films, reduced order leads to broadening of individual bands and increased decay rates compared to microcrystallites in powder. In particular, coupling to the Ag(111) single crystal increases the total decay rate up to three orders of magnitude to about 1 MHz, although we do not yet see indications for exciton quenching via charge transport. An envisioned molecular complex in close contact to two metallic leads is expected to decay at higher rates due to increased coupling, and molecular properties would need to be optimized accordingly. Quenching pathways in the kHz range such as coupling to vibrational modes, that determine the observed decay in powder or in solution, will be negligible, while molecular stability in direct vicinity of a metal and drastically reduced bleaching probability become essential. The observed efficient coupling to a metallic substrate is necessary to enhance the intrinsically low radiative rate of Eu$^{3+}$ emitters and paves the way to effective single-molecule junctions optimized for efficient outcoupling of the plasmonic excitations to the far field.
    
\section{Conflicts of interest}
There are no conflicts to declare.

\section{Acknowledgments}

We acknowledge funding by the Deutsche Forschungsgemeinschaft (DFG, German Research Foundation) through the Collaborative Research Center “4 f for Future” (CRC 1573, project number 471424360) project C1 and C2 and support by the Helmholtz Association via the programs Natural, Artificial, and Cognitive Information Processing (NACIP) and Materials Systems Engineering (MSE).

\bibliography{literature_arxiv.bib} 
\end{document}